\documentclass[a4paper] {article}
\usepackage{graphics,color,epsfig}
\def\linadj#1{\normalbaselines
	\multiply\lineskip#1 \divide\lineskip100
        \multiply\baselineskip#1 \divide\baselineskip100
	\multiply\lineskiplimit#1 \divide\lineskiplimit100 }

\begin{document}








\title{\bf  Effect of the curvature on a statistical model of Quark-Gluon-Plasma fireball in the hadronic medium }

\author{ S. Somorendro Singh, D. S. Gosain, Yogesh kumar \\and  Agam K. Jha$^*$ \footnote{Email:sssingh@physics.du.ac.in}}


\maketitle
\begin{center}
\Large   Department of Physics and Astrophysics, University of Delhi, Delhi - 110007, India \\   
\large $^*$Department of Physics, Sri Venkateswara College, University of Delhi, Delhi, India

\end{center}
\linadj{200}

\begin{abstract}
\large   The free energy of a Quark-Gluon Plasma fireball in the hadronic 
medium is calculated in the
Ramanathan et al. statistical model 
including the effect of curvature.
The result with this curvature is found to produce significant 
improvement from earlier results in all the parameters
we calculated. The 
 surface tension with
this curvature effect is found to be $~0.17~T_{c}^{3}$, which is two 
times the earlier value of surface tension which is $~0.078~T_{c}^{3} $, and 
it is nearly close
to the lattice value $0.24~T_{c}^{3}~ $. The speed of sound calculated
with curvature correction is still found to be smaller in comparision 
with the standard speed 
of sound in the QGP droplet.  \\

PACS number(s): 25.75.Ld, 12.38.MH, 21.65.+f


\end {abstract}

\vfill
\eject

\maketitle
 
\section{Introduction}

\large  Quantum chromodynamics predicts [1]a phase transition
from a deconfined to a confined matter of hadrons during the early
stages of the universe. Just in the beginning of the universe
 it was considered to be in the form of deconfined matter of 
quarks and gluons and subsequent process of cooling leads to
transformation into hadrons.
It is indeed a very complicated phenomena as predicted by the heavy-ion 
collider experiments. So, the formation of quark-gluon plasma 
droplet (fireball) is a very exciting field in the present day of
heavy ion collider physics [2].
 The central assumption in this approach treats 
the QGP-Hadron system as a quasi-static equilibrium 
enabling applicability of equilibrium statistical mechanics to 
this complicated system. 
This central assumption has been employed by many pioneers in this field
[2-3]. 

  As we know from the extensive literature [4], it is really needed to look at
the nucleation process produced by the statistical fluctuations of 
 the critical free energy difference between two phases. The 
model of Csernai-Kapusta et al. [3-5] uses the liquid drop model 
expansion for this nucleation. This model is used by Ramanathan et al. [4].
Now it is modified 
with the effect of curvature term in their free
energy of the liquid drop. This modified free energy is:

\begin{eqnarray} 
\Delta F = \frac {4\pi}{3} R^{3} [P_ {had}(T,\mu_{B}) - P_{q,g}(T, \mu_{B})] 
\nonumber \\
  + 4\pi R^{2} \sigma+ 8\pi C R                                                 \end{eqnarray}   

The first term represents the volume contribution, the second term is 
the surface contribution where $\sigma$ is the surface tension,
and the third term is the curvature term. We are interested
to see the effect of this curvature on the free energy. 
The critical radius $ R_{c}$ can be 
obtained by minimising eqn.(1) with respect to the droplet radius $R$, 
which gives two critical radii,

 \begin{equation}
 R_{c} = \frac {\sigma}{\Delta p}(1\pm \sqrt(1+\frac{2 \Delta p C}{\sigma^{2}}))
\end{equation}
with smaller radius corresponds to a local mimimum in free energy
and the larger to a local maximun. We take the local maximum value of the
radius as suitable case for our fireball system.  At the local
maxima, it has existence 
of stable solution. 
Then, we calculate
the surface tension through $\Delta F_{c}=F(R_{c},T)-F(0,T)$ and it
is obtained as:
\begin{equation}
  \sigma = \frac{2}{R_{c}}(\frac{3\Delta F_{c}}{8\pi R_{c}}-2C), 
\end{equation}
where ' C ' is the curvature coefficient.
To calculate the total relativistic density of states for quarks and gluons,
it is necessary to modify the density of states we 
obtained in our earlier paper~[4]. In that paper, we calculated 
the density of states adapting
the procedures of the Thomas-Fermi construction of the electronic density of
states for complex atoms and the Bethe density of states [6] for nucleons
in the complex nuclei as templates. To modify these total density of states,
we consider the higher order approximation scheme [7], which can give better
results in the calculation of total free energy of quarks and gluons. 
This higher 
order approximation is considered to be the curvature factor in this model.
In view of this situation, we construct density of states breifly 
highlighting the Ramanathan et al. statistical model in the hadronic medium
for obtaining free energy of quarks and gluons.
In the next section, we use total free energy to calculate 
the interfacial surface tension, thermodynamic variables and speed of sound
in this model with the curvature term.  
 \section{Detemination of density of states for the QGP 
droplet with curvature term}
To determine the density of states of QGP droplet,
the atomic model of large atomic number of Thomas and Fermi [6]is modified 
by Ramanathan et al. [4].
The total electronic density of states defined by the Thomas and Fermi 
in phase
space is:
\begin{equation}\label{3.11}
\int{\rho_{e}}(k)dk =  [ -2m V(k)]^{3/2} \nu / 3 \pi^2~.
\end{equation}

 or,

\begin{equation}\label{3.12}
\rho_e (k) = [\nu (2m)^{3/2} / 2 \pi^2]~ [-V(k)]^{1/2} \cdot \biggl[-\frac{dV(k)}{dk}\biggl]
\end{equation}

This model of electron is replaced by Ramanathan et al.'s 
model of QGP droplet with the corresponding density of
states for quark and gluon with a suitable 
QCD induced phenomenological potential $V (k) $. It is given as:

\begin{equation}
\int \rho_{q,g}dk=[-V_{conf}(k)]^{3}\nu/3\pi^{2}~,
\end{equation}
 
or,

\begin{equation}\label{3.13}
\rho_{q, g} (k) = (\nu / \pi^2) \lbrace (-V_{conf}(k))^{2}(-\frac{dV_{conf}(k)}{dk}) \rbrace_{q, g}, 
\end{equation}

where $\nu$ is the volume occupied by the QGP and $k$ is 
the relativistic four-momentum in natural units. $V_{conf}(k)$ could be 
any confining potential for quarks and gluons. This potential plays 
the role of a mean field potential in phase space similar to the 
mean field potential of the Thomas-Fermi scheme, but in a 
very different context-namely the QGP-Hadron system.   
The distinction of these two model is that Thomas-Fermi 
model deals about low temperature feature 
whereas QGP is  high
temperature feature which
subsequently leads to transformation into hadrons. Taking into account
of all these factors, the density of states was modified  with
a suitable parametrization factor in the dynamics of the QGP fluid. 
Thus, if we consider
that the curvature is essential higher 
order correction factor in free energy expansion, then 
the density of states is further
modified by using the technique of Neergaard
et al. [7],so that it gives more effective results in the free energy.
 So the modified density of states for quarks and gluons
is given as:
\begin{eqnarray}\label{3.15}
  \rho(k)=\rho_{q,g}(k)+\rho(k)_{cur}\nonumber \\
where,~ 
\label{3.15}  
  \rho(k)_{cur}=C_{i} \int ds (\frac{1}{R_{1}}+\frac{1}{R_{2}}) 
\end{eqnarray}
here,
$~C_{i}~$ is function of $k/m $ and depends on the type of field and 
on the boundary conditions. The different values of  $~C_{i}$ for different
particles are given as:
\begin{eqnarray}
   C_{s}=\frac{1}{12 \pi^{2}}[ 1-\frac{3 k}{2 m}(\frac{\pi}{2}-arctan(\frac{k}{m}))],
\end{eqnarray}
\begin{eqnarray}
   C_{g}= - \frac{1}{6 \pi^{2}}~ and~
\end{eqnarray}
\begin{eqnarray}
   C_{u}=C_{d}= -\frac{1}{ 24 \pi^{2}},
\end{eqnarray}
 which are in the limit of dynamical quark mass.
 
\section{Mean-field inter-quark potential and the Free energy}

To calculate the interacting potential within the system
, it is very ideal to find the effective mean field
potential among the quarks and quark-gluon. 
This effective potential is obtained through thermal hamiltonian
for the QGP [8].
\begin{equation}\label{3.18}
V_{\mbox{eff}}(k) = (1/2k)\gamma_{g,q} ~ g^{2} (k) T^{2}~.
\end{equation}
where,
 $g(k)$ for first order QCD running coupling constant, 
which for quarks with three flavors is, [8]

\begin{equation}\label{3.17}
g^2(k) =(4/3)(12\pi/27)\lbrace 1/ \ln(1+k^{2}/\Lambda^{2}) \rbrace ~,
\end{equation}

 In the above expression, $\Lambda = 150 MeV$ is QCD parameter
with our parametrization factors $\gamma_{g,q}$ which we take 
$\gamma_{q}=1/6$~and~$\gamma_{g}=~8~$~or~$ 6~$ times  $\gamma_{q}$. This value fits 
the lattice QCD simulation[9].
This effective perturbed potential 
will have minimum value 
at each point of phase space. That is :

\begin{equation}\label{3.19}
V(k_{min})=(\gamma_{g,q}N^{\frac{1}{3}} T^{2} \Lambda^2 / 2)^{1/4},
\end{equation}

where $N=(4/3 )(12 \pi / 27)$.  

So, this is called as low energy cut off in 
the model leading to finite integrals by avoiding 
the infra-red divergence. It is of the same order 
of magnitude as $\Lambda$ and 
$T$. We calculate the free energy for quarks and gluons
with the above modified density of states. 
\begin{equation}\label{3.20}
F_i = \mp T g_i \int dk \rho_i (k) \ln (1 \pm e^{-(\sqrt{m_{i}^2 + k^2}) /T})~,
\end{equation}

 where $\rho_{i}(k)$ is the modified density of states of the 
particular particle $i$ (quarks, gluons, pions etc.) 
and $g_{i}$ is the 
degeneracy factor (color and particle-antiparticle degeneracy) 
which is $6$ for 
quarks and $8$ for gluons [7]and $3$ for pions.

The density of state for interfacial surface is calculated through
the Weyl's model and Ramanathan et al's model. It is obtained as:
\begin{equation}
  \rho_{interface}(k)= \frac{4 \pi R^{2} k^{2}}{16 \pi}
\end{equation}  
Therefore, the interfacial energy
obtained through a scalar 
Weyl-surface in Ramanathan et.al[4,10-11] with  suitable 
modification to take care of 
the hydrodynamic effects [12] is:  
\begin{equation}\label{3.22}
  F_{interface}= \frac{1}{4}\gamma R^{2}T^{3} 
\end{equation}
where
\begin{equation}\label{3.22}
\gamma = \sqrt{2}\times \sqrt{(1/\gamma_{g})^{2} + (1 / \gamma_{q})^{2}},
\end{equation}

which is the effective rms value of the flow parameter of the quarks
and gluons respectively.
The pion free energy is [10]

\begin{equation}\label{3.25}
F_{\pi} = (3T/2\pi^2 )\nu \int_0^{\infty} k^2 dk \ln (1 - e^{-\sqrt{m_{\pi}^2 + k^2} / T})~.
\end{equation}

 For the quark masses we use the current (dynamic) quark masses $m_0 = m_d = 0 ~ MeV$ and $m_s = 150 ~ MeV$, just as 
in reference[10].

We can thus compute the total free energy $F_{total}$ as,
\begin{equation}
 F_{total}=\sum_{i} F_{i}~+~F_{interface}~+~F_{\pi},
\end{equation}
~ where $i$ stands for $u$,$d$
and $s$~quark and gluon.   
\section{Interfacial surface tension}
    
From the above equation (20),
we can compute the free-energy change with respect to the 
droplet radius for two promising 
parametrisation factors. As the value of these
two cases give more exciting
results in the free energy shown in Fig. $1$ and Fig.$2$, we 
expect that the formation
of droplet with observable critical droplet radius
will exhibit the proper transition features in the band of
temperature $ 150~ MeV$ to $170~MeV$ which are expected from
lattice calculations too.
 
\begin{figure}
\begin{center}
\epsfig{figure=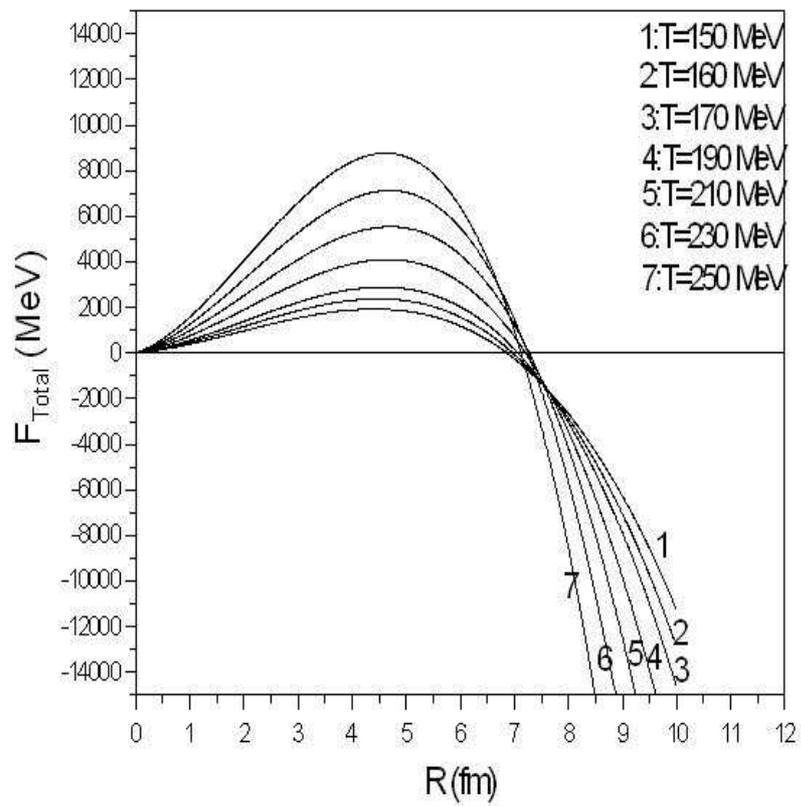,height=5.5in,width=4.5in}
\label{fig4}
\caption{\large $F_{total}$ at $\gamma_{g} = 6\gamma_{q}$, $ \gamma_{q} = 1/6 $ for various temperatures.}
\end{center}
\end{figure}

\begin{figure}
\begin{center}
\epsfig{figure=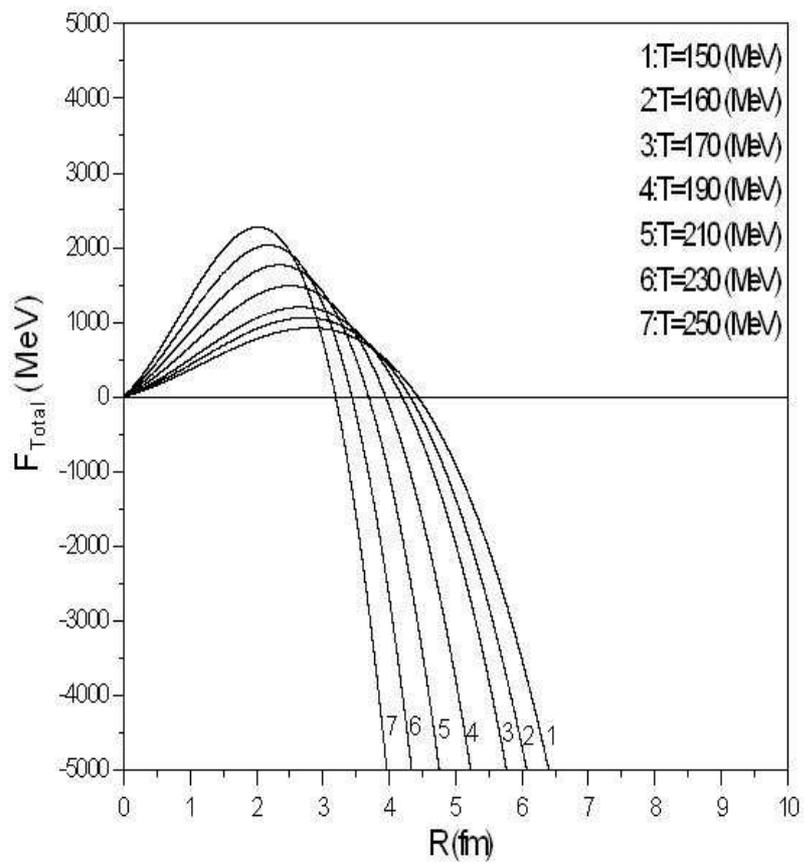,height=5.5in,width=4.5in}
\label{fig5}
\caption{\large $F_{total}$ at $\gamma_{g} = 8\gamma_{q}$ , $ \gamma_{q} = 1/6 $ for various temperatures.}
\end{center}
\end{figure}
\vfill
\eject
The Figs. $1$ and $2$ are the two most promising scenarios of our 
model which exhibit measurable droplet radius of the order 
of few fermi, and show significant barrier amplitudes in free 
energy which account for the nucleation rate of droplet formation, which 
was shown in the earlier paper too$[5]$. So, we use these promising values
of the free energy with curvature and look at the changes
produced by this effect.  
The set of parameters leading to fig $1$ seems more realistic because of 
stability of droplet size at free energy nearly
at critical radius irrespective of the transition 
temperature unlike in fig. $2$.   
From the values of the critical free-energy at the 
corresponding critical fireball radius that can be extracted 
from the Figs. 1 and 2, we can compute the surface tension of 
the fireball using equation (3) as listed in Tables 1 and 2.  

\begin{quote}
\begin{tabular}{|r|r|r|r|r|}
\hline
$T_{c}$&$\Delta F_{c}$&$R_{c}$&$\sigma$&$\frac{\sigma}{T_{c}^{3}}$\\  
$(MeV)$&$(MeV)$&$(fm)$&$(MeV/fm^{2})$&\\
\hline
150&921.93&2.814&14.798&0.17\\
160&1059.00&2.756&17.949&0.17\\
170&1200.00&2.665&21.528&0.17\\
190&1486.00&2.506&30.083&0.17\\
210&1765.00&2.344&40.586&0.17\\
230&2030.00&2.182&53.365&0.17\\
250&2272.00&2.024&68.540&0.17\\
\hline
\end{tabular}
\end{quote}
Table-1 for Surface Tension of QGP droplet at $\gamma_{g}=8\gamma_{q}$,$ \gamma_{q} = 1/6$.
\vfill

\begin{quote}
\begin{tabular}{|r|r|r|r|r|}
\hline
$T_{c}$&$\Delta F_{c}$&$R_{c}$&$\sigma$&$\frac{\sigma}{T_{c}^{3}}$\\
$(MeV)$&$(MeV)$&$(fm)$&$(MeV/fm^{2})$&\\
\hline
150&1928.00&4.416&9.646&0.17\\
160&2377.00&4.488&18.069&0.17\\
170&2890.00&4.555&21.667&0.17\\
190&4104.00&4.659&30.233&0.17\\                                               
210&5541.00&4.713&40.735&0.17\\                                                 
230&7132.00&4.703&53.461&0.17\\
250&8775.00&4.618&68.974&0.17\\
\hline
\end{tabular}                                       
\end{quote}
Table-2 for Surface Tension of QGP droplet at $\gamma_{g}=6\gamma_{q}$,$ \gamma_{q} = 1/6$. 
 
The constant result of $\sigma/T_{c}^{3}$ shows that the surface tension
is independent of transition
temperature and the values of the parametrisation factor.

\vfill
\eject 

\section{Thermodynamic variables and nature of the phase transition}

The thermodynamic properties give the following standard entities
such as:

\begin{equation}
   S =-({\partial{F}}/{\partial{T}})~.
\end{equation}

\begin{equation}
   C_{V}=T( {\partial{S}}/{\partial{T}})_{V}~.
\end{equation}

\begin{equation}
   C_{S}^{2}={S}/{C_{V}}
\end{equation} 

These entities can be obtained from the total
free energy for these two promising parametrisation
factors $\gamma_{g}= 6 \gamma_{q}~ and~ \gamma_{g}= 8 \gamma_{q} $.
The behaviour of entropy, $S$ vs temperature, $T$ and the heat capacity  
at constant volume, $C_{V}$
 with 
temperature, $T$ too indicate the real nature of the phase 
transition of the system. These output characteristics are shown 
in the following Figs. $3-8$.

\begin{figure}
\epsfig{figure=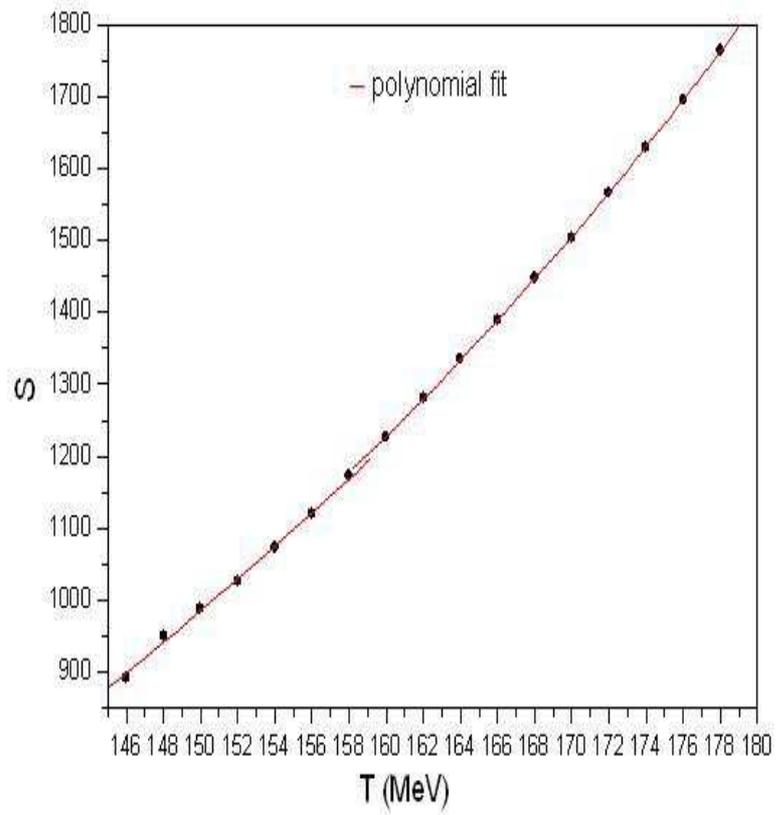,height=5.5in,width=4.5in}
\label{st_6.eps}
\caption{\large  Variation of $S$ with temperature T at $\gamma_{g} = 6\gamma_{q}$ , $ \gamma_{q} = 1/6 $.}
\end{figure}
 \begin{figure}
\epsfig{figure=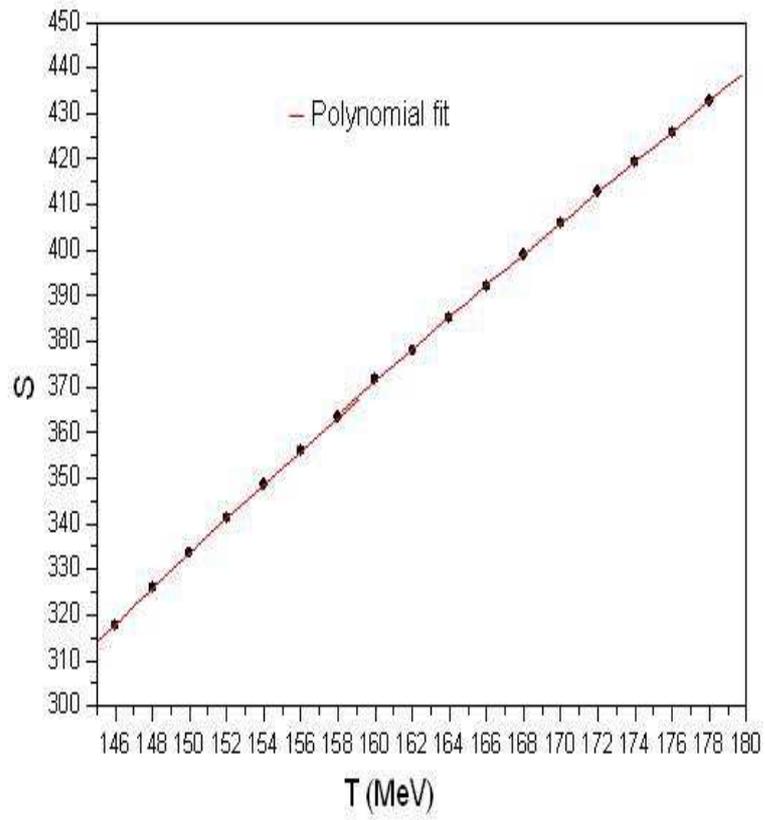,height=5.5in,width=4.5in}
\label{st_8.eps}
\caption{\large  Variation of $S$ with temperature T at $\gamma_{g} = 8\gamma_{q}$ , $ \gamma_{q} = 1/6 $.}
\end{figure}
\vfill
\eject
The above figs. $3$ and $4$ indicate that there exists a very 
weak discontinuity in the vicinity of $T_{c}=160~MeV$, in the 
entropy even though the curvature term
is included in the total free energy with parametrisation
 factor $\gamma_{g}= 6 \gamma_{q}$ and a very mild discontinous in
the entropy of the parametrisation factor $\gamma_{g}= 8 \gamma_{q}$. It 
is happened 
in the first order thermodynamic variable. The discontinuity is 
just of the order of one standard deviation of the entropy variable and 
therefore indeed claim to be a very weak transition . For the second 
order variable $C_{V}$ shown in figs. $5$ and $6$, there is absolutely 
no discontinuity
in the variable with increase in temperature.  
   
 \begin{figure}
\epsfig{figure=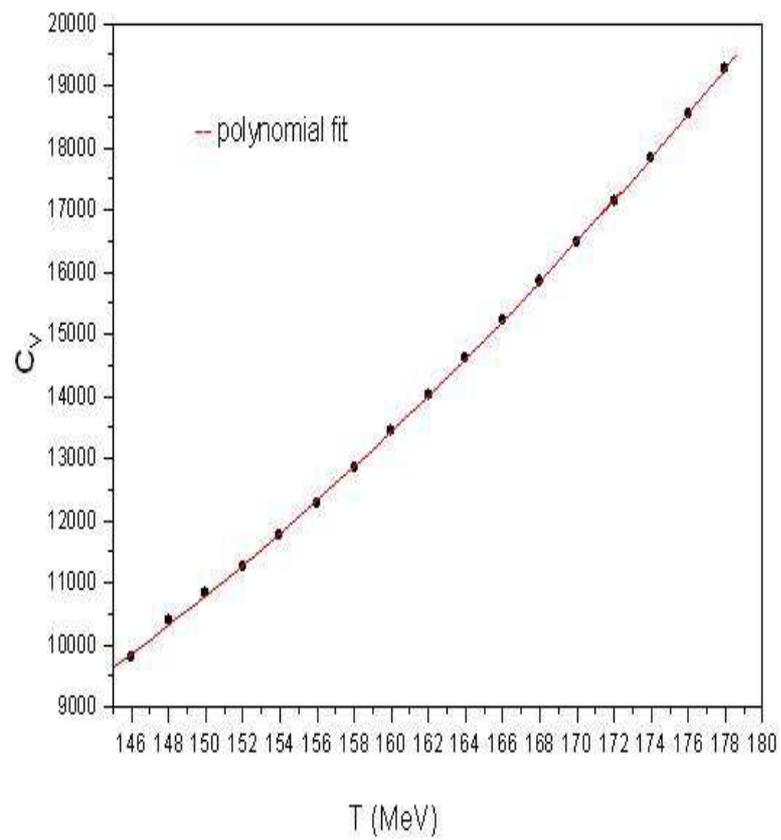,height=5.5in,width=4.5in}
\label{Cv_6.eps}
\caption{\large Variation of specific heat $C_{V}$ with temperature T at $\gamma_{g} = 6\gamma_{q}$ , $ \gamma_{q} = 1/6 $.}
\end{figure}

\begin{figure}
\epsfig{figure=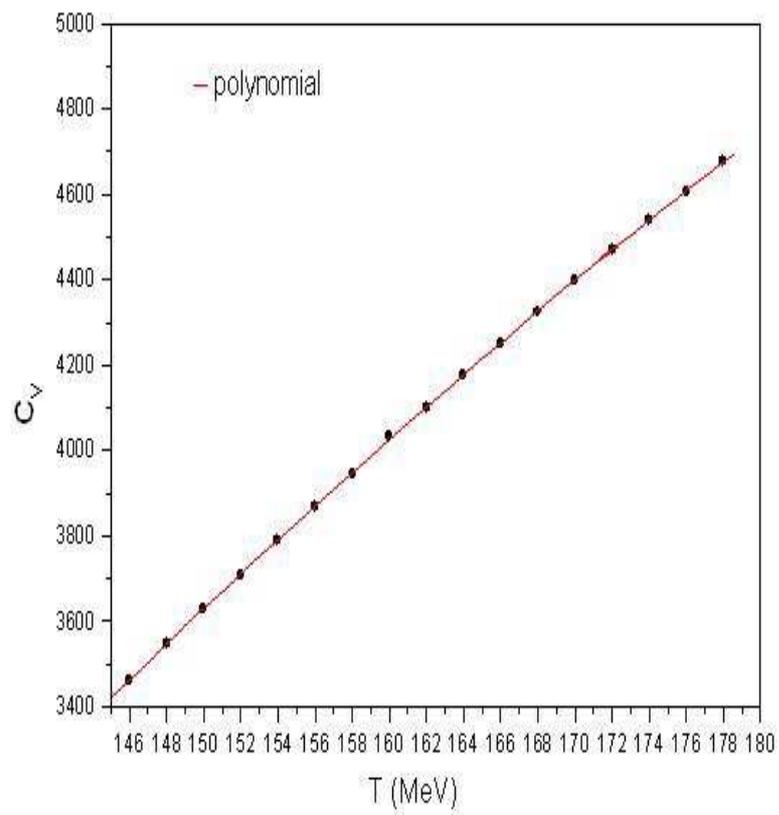,height=5.5in,width=4.5in}
\label{Cv_8.eps}
\caption{\large Variation of specific heat $C_{V}$ with temperature T at $\gamma_{g} = 8\gamma_{q}$ , $ \gamma_{q} = 1/6 $.}
\end{figure}
\vfill
\eject
Even though the curvature term is accounted for our calculations
there is a little improvement in the speed of sound.
The result is rather low in comparision with the lattice 
calculations of speed of sound. The model calculation of sound speed
is also independent of the transition temperature and model 
parametrization factor. 

\begin{figure}
\epsfig{figure=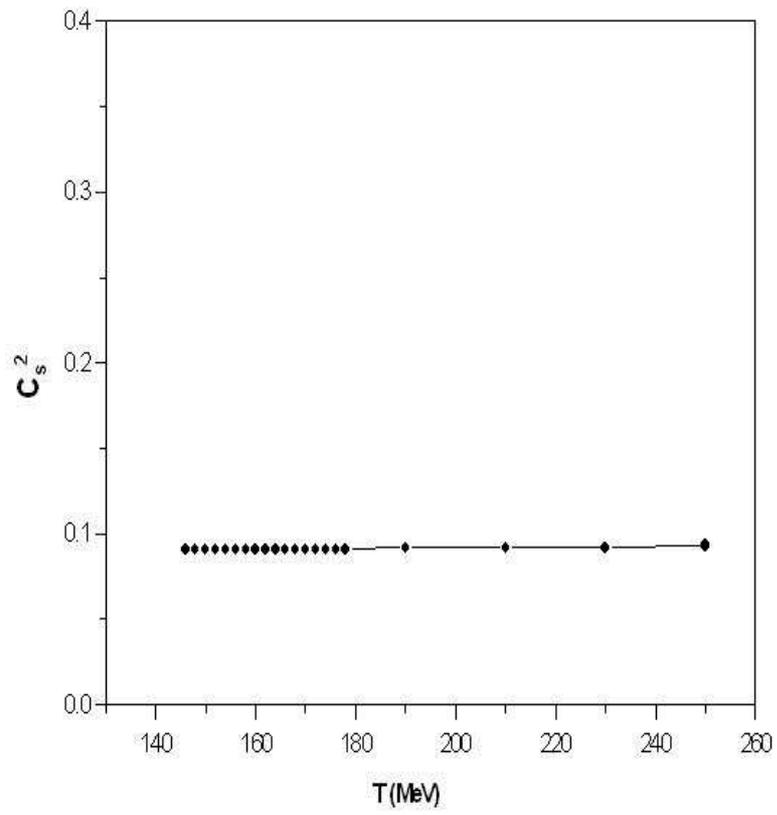,height=5.5in,width=4.5in}
\label{tCs6.eps}
\caption{\large Variation of speed of sound squared $ C_{S}^{2}$ with temperature T at $\gamma_{g} = 6\gamma_{q}$ , $ \gamma_{q} = 1/6 $.}
\end{figure}

\begin{figure}
\epsfig{figure=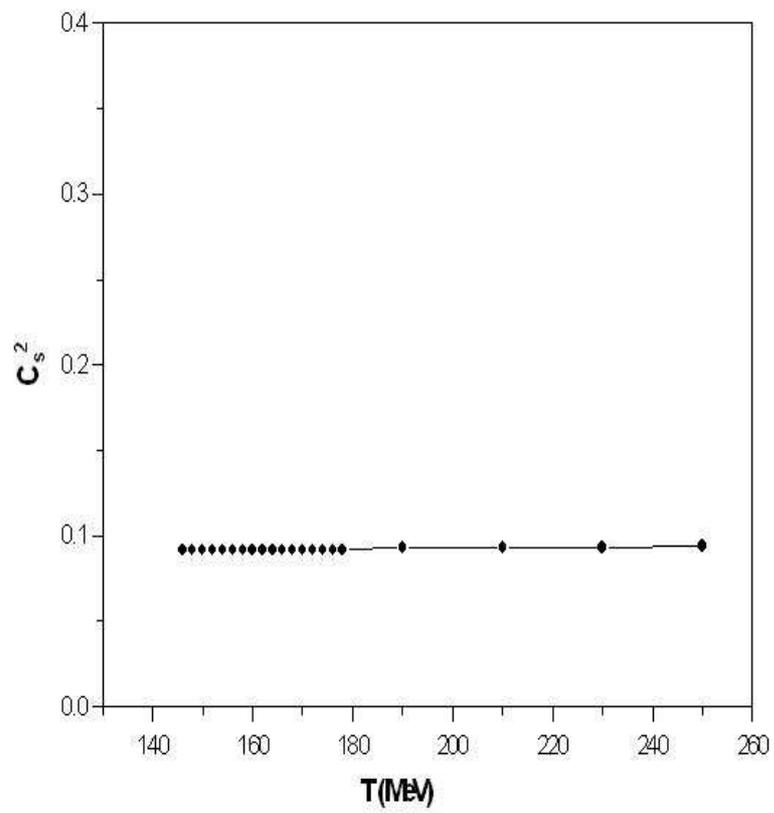,height=5.5in,width=4.5in}
\label{tCs8.eps}
\caption{\large Variation of speed of sound squared $C_{S}^{2}$ with temperature T at $\gamma_{g} = 8\gamma_{q}$ , $ \gamma_{q} = 1/6 $.}
\end{figure}
\vfill
\eject

\section{Conclusion:}
The free energy calculations for the two promising parametrization factors
are shown in the Figs.[1] and [2]. There is good improvement in the free 
energy with this curvature correction. The energy at $ R_{c}$ is
about two times the energy without the curvature and there is increase in
the maximum stable droplet size. Due to this significant effect in the
amplitude of the free energy, we determine the surface tension,
thermodynamic variables 
and speed of
sound in QGP for these two factors. There is significant 
increase in surface tension with the critical temperature in this modified
free energy. This shows good improvement over the earlier results of 
Ramanathan et al. without 
curvature. This leading order term of curvature has good impact to find the 
surface tension. The calculated value of  
surface tension is found to be $0.17~T_{c}^{3}$, which is two times
the earlier results that is $ 0.078~T_{c}^{3}$  [5] and 
this present value is nearly close to higher value of lattice results 
that is$~ \leq 0.24 T_{c}^{3}$[7,13]. The result is still consistent 
with earlier ones in terms of its constancy through out the temperature
and parametrization factors.
It is perfect conformity with the latest QCD simulations [9].
The model still shows weakly first order phase transition 
at the temperature in the range $(160~\pm~5)~MeV$ with 
this effect of curvature in the free energy
and it is expected with the current feature of QGP-Hadron 
phase transition too [1]. Thus, 
this transition is a mild characteristic in nature and 
 discontinuity is found
in the first order thermodynamic variable 
namely the entropy $S$ as shown in Figs.$3$ and $4$.
Lastly, the speed of sound with curvature correction
is  compared with the earlier result without curvature.
The improvement in the speed of sound is very small
. We obtained this value
earlier nearly close to $~0.09~T/T_{c}$. Presently it is found 
to be $~0.1~T/T_{c}$ 
but still the result is 
showing low value with the lattice QCD as before[13]. This low 
result is expected
with the recent model calculations [14]and lattice simulations that include
dynamical quarks
[15], but as should be expected, it is at variance with pure 
gauge lattice results[16].\\   
\\
{\bf Acknowledgments:}
\\
\\
The authors (D.S.G. and Y.K.) are  
highly obliged to the department for giving them the opportunity
to do research here and very thankful to CSIR and Rajiv Gandhi Fellowship
for financial support.
\\
\\  
{\bf References :}
\\
\\
\begin{enumerate}
\item{C. Y. Wong, Introduction to High -Energy Heavy Ion Collisions (World Scientific, Singapore,1994); L. P. Csernai, Introduction to Relativistic Heavy-Ion Collisions (Wiley, New York,1994); R. C. Hwa, Quark-Gluon Plasma (World Scientific, Singapore, 1990)}.
\item{H. Satz, CERN-TH-2590, 18pp (1978); F.Karsch, E. Laer
mann, A. Peikert, Ch. Schmidt and S. Stickan, Nucl. Phys. B
94, 411 (2001); F. Karsch and H. Satz, Nucl. Phys. A 702, 373(2002).}
\item{L. P. Csernai, J. I. Kapusta, E. Osnes, Phys. Rev. D 67,
045003 (2003); J. I. Kapusta, R. Venugopal, A. P. Vischer, Phys. Rev. C51,
901(1995); L. P. Csernai, J. I. Kapusta, Phys. Rev. Lett.69, 737(1992);
L. P. Csernai, J. I. Kapusta, Gy. Kluge, E. E. Zabrodin, Z. Phys. C 58, 453(1993).}
\item{R. Ramanathan, Y. K. Mathur, K. K. Gupta and Agam K. Jha, Phys. Rev. C\textbf{70}, 027903 (2004); R. Ramanathan, K. K. Gupta, Agam K. Jha and S. S. Singh, Pram. J. Phys. 757(2007).}
\item{  B. D. Malhotra and R. Ramanathan, Phys. Lett. A \textbf{108}, 153 (1985); See also in references (3).}
\item{A. D. Linde, Nucl. Phys. B \textbf{216}, 421 (1983);  E. Fermi, Zeit F. Physik \textbf{48}, 73 (1928); L. H. Thomas , Proc. camb. Phil. Soc.\textbf{23}, 542 (1927); H. A. Bethe, Rev. Mod. Phys.\textbf{9}, 69 (1937).}
\item{ G. Neergaad and
J. Madsen, Phys. Rev. D60, 054011(1999); M. B. Christiansen and J. Madsen,
J. Phys. G 23, 2039(1997).}
\item{ A. Peshier, B. Kampfer, O. P. Pavlenko and G. Soff, Phys. Lett. B \textbf{337}, 235 (1994); V. Goloviznin and H. Satz, Z. Phys. C \textbf{57}, 671 (1993).}
\item{Y. Iwasaki, K. Kanaya, L. K\"{a}rkk\"{a}inen, K. Rummukainen and
T. Yoshic, Phys. Rev. D49, 3540(1994)
.}
\item{I. Mardor and B. Svetitsky, Phys. Rev. D 44, 878(1991); M. G. Mustafa, A. Ansari, Z. Phys. C 57, 51(1993); M. G. Mustafa, D. K. Srivastava and B. Sinha, Euro.Phys. J. C5, 711(1998).}
\item{H. Weyl, Nachr. Akad. Wiss Gottingen 110 (1911).}
\item{J. Solfrank, P. Husvinen, M. Kataja, P.V Ruuskanen, M. Prakash
and R. venugopalan, Phys. Rev. C 55, 392(1997); C. M. Hung and E. Shuryak,
Phys. Rev. C 57, 1891(1998); E. Shuryak, Prog. Part. Nucl. Phys. 53, 273(2004).}
\item{Rajiv V. Gavai, J.O. Phys.: Cofer. Series 50, 87(2006);
K. M. Udayananda, P. Sethumadhavan and V. M. Bannur, Phys. Rev. C76, 044908(2007).}
\item{Sanjay K. Ghosh, Tamal K. Mukherjee, Munshi G. Mustafa, and Rajarshi Ray, Phys. Rev. D\textbf{73}, 114007(2006).}
\item{A. Ali Khan et al., Phys. Rev. D \textbf{64}, 074510(2001); Yasumichi Aoki, Zoltan Fodor, Sandor D. Katz and Kalman K. Szabo, JHEP \textbf{01}, 089(2006).}
\item{ Rajiv V. Gavai, Sourendu Gupta, Swagato Mukherjee, hep-lat/0506015;
.}   
\end{enumerate}

\end{document}